\documentclass[11pt]{article}
\usepackage{amssymb,amsmath,amsfonts}
\usepackage{graphicx}
\usepackage{graphics}
\usepackage{eepic,epsfig}

\begin{document}

\title{ANGULAR DISTRIBUTION OF PHOTONS IN COHERENT BREMSSTRAHLUNG IN
DEFORMED CRYSTALS}
\author{V. V. Parazian \\
\textit{Institute of Applied Problems in Physics,}\\
\textit{25 Nersessian Str., 0014 Yerevan, Armenia}}
\maketitle

\begin{abstract}
We investigate the angular distribution of photons in the coherent
bremsstrahlung process by high-energy electrons in a periodically deformed
single crystal with a complex base. The formula for the corresponding
differential cross-section is derived for an arbitrary deformation field.
The case is considered in detail when the electron enters into the crystal
at small angles with respect to a crystallographic axis. The results of the
numerical calculations are presented for ${\mathrm{SiO}}_{2}$ single crystal
and Moliere parameterization of the screened atomic potentials in the case
of the deformation field generated by the acoustic wave of S -type.
\end{abstract}

\bigskip


\bigskip


\bigskip

\section{Introduction}

The investigation of high-energy electromagnetic processes in crystals is of
interest not only from the viewpoint of underlying physics but also from the
viewpoint of practical applications. From the point of view of controlling
the parameters of various processes in a medium, it is of interest to
investigate the influence of external fields, such as acoustic waves,
temperature gradient etc., on the corresponding characteristics. The
considerations of concrete processes, such as diffraction radiation \cite%
{MkrtchLSh}, transition radiation \cite{LShAHMkrtch}, parametric X-radiation
\cite{MkrtchAsl}, channelling radiation \cite{MkrtchGaspar}, bremsstrahlung
by high-energy electrons \cite{Paraz19}, have shown that the external fields
can essentially change the angular-frequency characteristics of the
radiation intensities. Recently there has been broad interest to compact
crystalline undulators with periodically deformed crystallographic planes an
efficient source of high energy photons \cite{Koro97} (for a review with
more complete list of references see \cite{Koro04}). The coherent
bremsstrahlung of high-energy electrons moving in a crystal is one of the
most effective methods to produce intense beams of highly polarized and
monochromatic photons. Such radiation has a number of remarkable properties
and at present it has found many important applications. This motivates the
importance of investigations for various mechanisms of controlling the
radiation parameters. In \cite{Paraz19}\ we have discussed the influence of
hypersonic waves excited in a crystal on the process of the bremsstrahlung
of high-energy electrons. To have an essential influence of the acoustic
wave high-frequency hypersound is needed. Usually this type of waves is
excited by high-frequency electromagnetic field through the piezoelectric
effect in crystals with a complex base. In the paper \cite{Parazbrem}, we
have generalized the results of \cite{Paraz19} for crystals with a complex
base and for acoustic waves with an arbitrary profile. For the experimental
detection of final particles in the process of coherent bremsstrahlung it is
important to know their angular distribution. In the present paper the
angular distribution of photons in the coherent bremsstrahlung in crystals
is investigated in the presence of hypersonic wave. The numerical
calculations are carried out for the quartz single crystal and for the
electrons of energy 20 GeV.

The paper is organized as follows. In the next section we derive the general
formula for the coherent part of the bremsstrahlung cross-section averaged
on thermal fluctuations and the conditions are specified under which the
influence of the deformation field can be considerable. In Sec. 3 the
analysis of the general formula is presanted in the cases when the electron
enters into the crystal at small angles with respect to crystallographic
axes or planes and the results of the numerical calculations for the
cross-section as a function of the angle between the momenta of electron and
photon. Sec. 4 summarizes the main results of the paper.

\section{Angular dependence of the cross-section}

\label{sec2:form}

We consider the bremsstrahlung by high-energy electrons in a single crystal.
The corresponding cross section can be obtained from the cross section for
the bremsstrahlung on a single atom derived in \cite{AkhBoldShul}. The
formulae \cite{BetheHeitler} practically cannot be used for the
investigation of bremsstrahlung in case of potentials of a complex form, for
example in a crystal, as errors of approximations, admissible in the case of
a single scattering center, can become inadmissible greater for many
centers. By making use of the results derived in \cite{AkhBoldShul}, the
differential cross section for the bremsstrahlung on a single atom is
presented in the form (the system of units $\hbar =c=1$ is used)%
\begin{eqnarray}
\frac{d^{5}\sigma _{0b}}{d\omega dq_{\Vert }d\mathbf{q}_{\bot }dy} &=&\frac{%
e^{2}}{8\pi ^{4}\epsilon _{1}^{2}}\frac{q_{\bot }^{2}}{q_{\Vert }^{2}}\left[
1+\frac{\omega ^{2}}{2\epsilon _{1}\epsilon _{2}}-4y^{2}\frac{\delta }{%
q_{\Vert }}\left( 1-\frac{\delta }{q_{\Vert }}\right) \right] \frac{%
\left\vert u\left( \mathbf{q}\right) \right\vert ^{2}}{\sqrt{1-y^{2}}}
\notag \\
&=&\left\vert u\left( \mathbf{q}\right) \right\vert ^{2}\sigma _{0}\left(
\mathbf{q,}y\right) ,  \label{sigindatom}
\end{eqnarray}%
where $e$ is the electron charge, $\omega $, $\epsilon _{1}$, $\epsilon _{2}$%
, are the energies of photon, initial and final electrons respectively, $%
\delta =m_{e}^{2}\omega /\left( 2\epsilon _{1}\epsilon _{2}\right) $, $m_{e}$
is the mass of electron, $q_{\Vert }$ and $q_{\bot }$ are the components of
the vector of momentum transfer $\mathbf{q}$, $\mathbf{q=p}_{1}-\mathbf{p}%
_{2}-\mathbf{k}$ ($\mathbf{k}$, $\mathbf{p}_{1}$, $\mathbf{p}_{2}$ are the
momenta of photon, initial and final electrons respectively), parallel and
perpendicular to the direction of the electron momentum, $u\left( \mathbf{q}%
\right) $ is the Fourier transform of the atom potential. The variable $y$
is expressed in terms of the angle $\theta _{\gamma }$ between the momenta $%
\mathbf{p}_{1}$ and $\mathbf{k}$ by the following relation:
\begin{equation}
\left( \frac{\epsilon _{1}\theta _{\gamma }}{m_{e}}\right) ^{2}=\frac{1}{%
\delta }\left( q_{\Vert }-\delta -\frac{q_{\bot }^{2}}{2\epsilon _{1}}+\frac{%
q_{\bot }^{2}\delta }{m_{e}^{2}}\right) +y\frac{2q_{\bot }}{m_{e}}\left(
\frac{q_{\Vert }}{\delta }-1-\frac{q_{\bot }^{2}}{2\epsilon _{1}\delta }%
\right) ^{\frac{1}{2}}.  \label{tetplyshul}
\end{equation}%
The regions of variables $q_{\parallel }$, $q_{\perp }$, $y$ in
cross-section (\ref{sigindatom}) are as follows \cite{AkhBoldShul}:
\begin{equation}
q_{\parallel }\geqslant \delta +\frac{q_{\perp }^{2}}{2\omega }%
,\;-1\leqslant y\leqslant 1,\;q_{\perp }\geqslant 0.  \label{paymanyq}
\end{equation}

The differential cross-section for the bremsstrahlung in a crystal can be
written in the form \cite{Parazbrem}

\begin{equation}
\mathbf{\sigma }\left( \mathbf{q,}y\right) \equiv \frac{d^{5}\sigma _{0b}}{%
d\omega dq_{\parallel }d\mathbf{q}_{\perp }dy}=\left\vert \sum_{n,j}u_{%
\mathbf{q}}^{\left( j\right) }e^{i\mathbf{qr}_{n}^{\left( j\right)
}}\right\vert ^{2}\sigma _{0}\left( \mathbf{q,}y\right) ,  \label{sigkrysdef}
\end{equation}%
where $\mathbf{r}_{n}^{\left( j\right) }$ is the position of an atom in the
crystal. In the discussion that follows, the collective index $n$ enumerates
the elementary cell and the subscript $j$ enumerates the atoms in a given
cell of a crystal. Here $\mathbf{q}$ is the momentum transferred to the
crystal, $\mathbf{q=p}_{1}-\mathbf{p}_{2}-\mathbf{k}$ and the differential
cross-section in a crystal given by (\ref{sigkrysdef}), differs from the
cross-section on an isolated atom by the interference factor which is
responsible for coherent effects arising due to the periodical arrangement
of the atoms in the crystal.

After averaging on thermal fluctuations, the cross-section is written in the
form (see, for instance \cite{Parazbrem} for the case of a crystal with a
simple cell)%
\begin{equation}
\sigma \left( \mathbf{q,}y\right) =\left\{ N\sum_{j}\left\vert u_{\mathbf{q}%
}^{\left( j\right) }\right\vert ^{2}\left( 1-e^{-q^{2}\overline{%
u_{t}^{\left( j\right) 2}}}\right) +\left\vert \sum_{n,j}u_{\mathbf{q}%
}^{\left( j\right) }e^{i\mathbf{qr}_{n0}^{\left( j\right) }}e^{-\frac{1}{2}%
q^{2}\overline{u_{t}^{\left( j\right) 2}}}\right\vert ^{2}\right\} \sigma
_{0}\left( \mathbf{q,}y\right) ,  \label{sigtermik}
\end{equation}%
where $N$ is the number of cells, $\mathbf{r}_{n0}^{\left( j\right) }$
determine the equilibrium positions of the atoms, $\overline{u_{t}^{\left(
j\right) 2}}$ is the temperature-dependent mean-squared amplitude of the
thermal vibrations of the $j$th atom, $e^{-q^{2}\overline{u_{t}^{\left(
j\right) 2}}}$ is the corresponding Debye-Waller factor. In formula (\ref%
{sigtermik}) the first term in figure braces does not depend on the
direction of the vector $\mathbf{k}$ and determines the contribution of
incoherent effects. The contribution of coherent effects is presented by the
second term. By taking into account the formula (\ref{sigindatom}) for the
cross-section on a single atom, in the region $\epsilon _{1}q_{\perp
}^{2}/\omega m_{e}^{2}\ll 1$ the corresponding part of the cross-section in
a crystal can be presented in the form
\begin{equation}
\sigma _{c}=\frac{e^{2}}{8\pi ^{4}\epsilon _{1}^{2}}\frac{q_{\perp }^{2}}{%
q_{\parallel }^{2}}\left[ 1+\frac{\omega ^{2}}{2\epsilon _{1}\epsilon _{2}}%
-4y^{2}\frac{\delta }{q_{\Vert }}\left( 1-\frac{\delta }{q_{\Vert }}\right) %
\right] \frac{1}{\sqrt{1-y^{2}}}\left\vert \sum_{n,j}u_{\mathbf{q}}^{\left(
j\right) }e^{i\mathbf{qr}_{n0}^{\left( j\right) }}e^{-\frac{1}{2}q^{2}%
\overline{u_{t}^{\left( j\right) 2}}}\right\vert ^{2}  \label{sigccrysfac}
\end{equation}

When external influences are present (for example, in the form of acoustic
waves) the positions of atoms in the crystal can be written as $\mathbf{r}%
_{n0}^{\left( j\right) }=\mathbf{r}_{ne}^{\left( j\right) }+\mathbf{u}%
_{n}^{\left( j\right) }$, where $\mathbf{r}_{ne}^{\left( j\right) }$
determines the equilibrium position of an atom in the situation without
deformation, $\mathbf{u}_{n}^{\left( j\right) }$ is the displacement of the
atom caused by the external influence. We consider deformations with the
periodical structure:
\begin{equation}
\mathbf{u}_{n}^{\left( j\right) }=\mathbf{u}_{0}f\left( \mathbf{k}_{s}%
\mathbf{r}_{ne}^{\left( j\right) }\right)  \label{deformperiod}
\end{equation}%
where $\mathbf{u}_{0}$ and $\mathbf{k}_{s}$\ are the amplitude and wave
vector corresponding to the deformation field, $f\left( x\right) $ is an
arbitrary function with the period $2\pi $, $\max $ $f\left( x\right) =1$.
In the discussion that folows, we assume that $f\left( x\right) \in
C^{\infty }\left( R\right) $. \ Further following \cite{Paraposdistr} and
having made similar steps we receive the following formula for the
differential cross section of coherent bremsstrahlung:%
\begin{equation}
\frac{d^{2}\sigma _{b}^{c}}{d\omega dy}=\frac{e^{2}N}{\pi \epsilon
_{1}^{2}N_{0}\Delta }\sum_{m,\mathbf{g}}\frac{g_{m\perp }^{2}}{g_{m\parallel
}^{2}}\left[ 1+\frac{\omega ^{2}}{2\epsilon _{1}\epsilon _{2}}-4y^{2}\frac{%
\delta }{g_{m\parallel }}\left( 1-\frac{\delta }{g_{m\parallel }}\right) %
\right] \frac{\left\vert F_{m}\left( \mathbf{g}_{m}\mathbf{u}_{0}\right)
\right\vert ^{2}\left\vert S\left( \mathbf{g}_{m},\mathbf{g}\right)
\right\vert ^{2}}{\sqrt{1-y^{2}}},  \label{sigcohgeneral}
\end{equation}%
where $N_{0}$ is the number of atoms in the crystal\ $\mathbf{g}_{m}=\mathbf{%
g-}m\mathbf{k}_{s}$, $\mathbf{g}$ is the reciprocal lattice vector, $-m%
\mathbf{k}_{s}$ stands for the momentum transfer to the external field, $%
g_{\Vert }$ and $g_{\bot }$ are the components of the vector $\mathbf{g}_{m}$%
. The function $F_{m}\left( x\right) $ is defined by relation%
\begin{equation}
F_{m}\left( x\right) =\frac{1}{2\pi }\int_{-\pi }^{\pi }e^{ixf\left(
t\right) -imt}dt,S\left( \mathbf{g,g}_{m}\right) =\sum_{j}u_{\mathbf{g}%
}^{\left( j\right) }e^{i\mathbf{g}_{m}\mathbf{\rho }^{\left( j\right) }}e^{-%
\frac{1}{2}g^{2}\overline{u_{t}^{\left( j\right) 2}}}  \label{Fmdefin}
\end{equation}
is the factor determined by the structure of the elementary cell. For a
lattice with a complex cell the coordinates of the atoms are written as $%
\mathbf{r}_{ne}^{\left( j\right) }=\mathbf{R}_{n}+\mathbf{\rho }_{j}$, with $%
\mathbf{R}_{n}$ being the positions of the atoms for one of primitive
lattices, and $\mathbf{\rho }_{j}$ are the equilibrium positions for other
atoms inside $n$-th elementary cell with respect to $\mathbf{R}_{n}$. Now
the relation between the variables $y$ and $\theta _{\gamma }$ is written in
the form:%
\begin{equation}
y=\frac{m_{e}}{2g_{m\perp }}\frac{\left( \epsilon _{1}\theta _{\gamma
}/m_{e}\right) ^{2}-1/\delta \left( g_{m\parallel }-\delta -g_{m\perp
}^{2}/\left( 2\epsilon _{1}\right) +g_{m\perp }^{2}\delta /m_{e}^{2}\right)
}{\left[ g_{m\parallel }/\delta -1-g_{m\perp }^{2}/\left( 2\epsilon
_{1}\delta \right) \right] ^{\frac{1}{2}}}  \label{ydefincrys}
\end{equation}

The regions of variables in cross-section (\ref{sigkrysdef}) are%
\begin{equation}
g_{m\parallel }\geq \delta +\frac{g_{m\perp }^{2}}{2\epsilon _{1}},\;-1\leq
y\leq 1,\;g_{m\perp }\geq 0.  \label{gmparalygmpercond}
\end{equation}%
For sinusoidal deformation field, $\ f\left( z\right) =\sin \left( z+\varphi
_{0}\right) $, one has the Fourier-transform%
\begin{equation}
F_{m}\left( x\right) =e^{im\varphi _{0}}J_{m}\left( x\right) ,
\label{FmBessel}
\end{equation}%
with the Bessel function $J_{m}\left( x\right) $.

The formula for the pair creation in an undeformed crystal is obtained from (%
\ref{sigcohgeneral}) taking $\mathbf{u}_{0}=0$. In this limit, the
contribution of the term with $m=0$ remains only with $F_{0}\left( 0\right)
=1$. Now we see that formula (\ref{sigcohgeneral}) differs from the formula
in an undeformed crystal by the replacement $\mathbf{g\rightarrow g}_{m}$,
and by the additional summation over $m$ with the weights $|F_{m}\left(
\mathbf{g}_{m}\mathbf{u}_{0}\right) |^{2}$. This corresponds to the presence
of an additional one-dimensional superlattice with the period $\lambda _{s}$
and the reciprocal lattice vector $m\mathbf{k}_{s}$, $m=0,\pm 1,\pm 2,...$ .
As the main contribution into the cross-section comes from the terms with $%
g_{m\parallel }\sim \delta $, the influence of the deformation field may be
considerable if $|mk_{s\parallel }|\gtrsim \delta $. Combining this with the
previous estimates, we find the following condition: $u_{0}/\lambda
_{s}\gtrsim a/4\pi ^{2}l_{c}$. At high energies one has $a/l_{c}\ll 1$ and
this condition can be consistent with the condition $u_{0}/\lambda _{s}\ll 1$%
.

In the presence of the deformation field the number of possibilities to
satisfy the condition $g_{m\parallel }\geq \delta +g_{m\perp }^{2}/\left(
2\epsilon _{1}\right) $ in the summation of formula (\ref{sigcohgeneral})
increases due to the term $mk_{s\parallel }$ in the expression for $%
g_{m\parallel }$. This leads to the appearance of additional peaks in the
angular distribution of the radiated positrons. After the integration of (%
\ref{sigcohgeneral}) over $y$, due to these additional peaks, there can be
an enhancement of the cross-section of the process \cite{Parazbrem}.

\section{Limiting cases and numerical results}

\label{sec3:an}

Now, we consider the most interesting case when the electron enters into the
crystal at small angle $\theta $\ with respect to the crystallographic $z$\
-axis of the orthogonal lattice. The corresponding reciprocal lattice vector
components are $g_{i}=2\pi n_{i}/a_{i},$ $n_{i}=0,\pm 1,\pm 2,,...$, where $%
a_{i},$ $i=1,2,3,$ are the lattice constants in the corresponding
directions. For the longitudinal component we can write
\begin{equation}
g_{m\parallel }=g_{mz}\cos \theta +\left( g_{my}\cos \alpha +g_{mx}\sin
\alpha \right) \sin \theta ,  \label{gmpargen}
\end{equation}%
where $\alpha $ is the angle between the projection of the vector $\mathbf{p}%
_{1}$ on the plane $(x,y)$ and axis $y$. For small angles $\theta $ the main
contribution into the cross-section comes from the summands with $g_{z}=0$.
Having made the replacement of variable $y\rightarrow \epsilon _{1}\theta
_{\gamma }/m_{e}$ using the formula (\ref{ydefincrys}) from formula (\ref%
{sigcohgeneral}) one finds%
\begin{eqnarray}
\frac{d^{2}\sigma _{b}^{c}}{d\omega d\left( \epsilon _{1}\theta _{\gamma
}/m_{e}\right) } &\approx &\frac{e^{2}N}{\pi \epsilon _{1}^{2}N_{0}\Delta }%
\sum_{m,g_{x},g_{y}}\frac{g_{\perp }^{2}}{g_{m\parallel }^{2}}\left[ 1+\frac{%
\omega ^{2}}{2\epsilon _{1}\epsilon _{2}}-4y^{2}\left( \theta _{\gamma
}\right) \frac{\delta }{g_{m\parallel }}\left( 1-\frac{\delta }{%
g_{m\parallel }}\right) \right]  \notag \\
&&\times \frac{\left\vert F_{m}\left( \mathbf{g}_{m}\mathbf{u}_{0}\right)
\right\vert ^{2}\left\vert S\left( \mathbf{g}_{m},\mathbf{g}\right)
\right\vert ^{2}}{\sqrt{1-y^{2}\left( \theta _{\gamma }\right) }}\frac{%
\epsilon _{1}\theta _{\gamma }/m_{e}}{\left( g_{\perp }/m_{e}\right) \left(
g_{m\parallel }/\delta -1-g_{\perp }^{2}/\left( 2\epsilon _{1}\delta \right)
\right) ^{\frac{1}{2}}},  \label{sigcasegz0}
\end{eqnarray}%
where the notation $y^{2}\left( \theta _{\gamma }\right) $ is intrduced in
accordance with:%
\begin{equation}
y^{2}\left( \theta _{\gamma }\right) =\frac{m_{e}^{2}}{4g_{\perp }^{2}}\frac{%
\left[ \left( \epsilon _{1}\theta _{\gamma }/m_{e}\right) ^{2}-\left(
1/\delta \right) \left( g_{m\parallel }-\delta -g_{\perp }^{2}/\left(
2\epsilon _{1}\right) +g_{\perp }^{2}\delta /m_{e}^{2}\right) \right] ^{2}}{%
g_{m\parallel }/\delta -1-g_{m\perp }^{2}/\left( 2\epsilon _{1}\delta
\right) }.  \label{y2casegz0}
\end{equation}%
In (\ref{sigcasegz0}) $g_{\perp }^{2}=g_{x}^{2}+g_{y}^{2}$, and the
summation goes over the region $g_{m\parallel }\geq \delta +g_{m\perp
}^{2}/\left( 2\epsilon _{1}\right) ,$ \ $0\leq y^{2}\left( \theta _{\gamma
}\right) \leq 1$ with
\begin{equation}
g_{m\parallel }\approx -mk_{z}+\left( g_{mx}\sin \alpha +g_{my}\cos \alpha
\right) \theta .  \label{gmparmain}
\end{equation}%
Note that in the argument of the functions $F_{m}$ and $S$ we have $\mathbf{g%
}_{m}\approx \left( g_{x},g_{y},0\right) $.

We now assume that the electron enters into the crystal at small angle $%
\theta $ with respect to the crystallographic axis $z$ and near the
crystallographic plane $\left( y,z\right) $ (the angle $\alpha $\ is small).
In this case with the change of $\delta $, the sum over $g_{x}$ and $g_{y}$
will drop sets of terms which lead to the abrupt change of the corresponding
cross-section. Two cases have to distinguish. Under the condition $\delta
\sim 2\pi \theta /a_{2}$, in Eq. (\ref{sigcasegz0}) for the longitudinal
component, one has
\begin{equation}
g_{m\parallel }\approx -mk_{s\parallel }+\theta g_{y}\geq \delta +\frac{%
g_{\perp }^{2}}{2\epsilon _{1}}.  \label{gmpartetgy}
\end{equation}

The formula (\ref{sigcasegz0}) can be further simplified under the
assumption $\mathbf{u}_{0}\perp \mathbf{a}_{1}$. In this case, in the
argument of the function $F_{m}$, one has $\mathbf{g}_{m}\mathbf{u}%
_{0}\approx g_{y}u_{0y}$ and we obtain the formula

\begin{eqnarray}
\frac{d^{2}\sigma _{b}^{c}}{d\omega d\left( \epsilon _{1}\theta _{\gamma
}/m_{e}\right) } &\approx &\frac{e^{2}N}{\pi ^{2}\epsilon
_{1}^{2}N_{0}\Delta }\sum_{m,g_{x},g_{y}}\frac{g_{\perp }^{2}}{g_{m\parallel
}^{2}}\left[ 1+\frac{\omega ^{2}}{2\epsilon _{1}\epsilon _{2}}-4y^{2}\left(
\theta _{\gamma }\right) \frac{\delta }{g_{m\parallel }}\left( 1-\frac{%
\delta }{g_{m\parallel }}\right) \right]  \notag \\
&&\frac{\left\vert F_{m}\left( g_{y}u_{y0}\right) \right\vert ^{2}\left\vert
S\left( \mathbf{g}_{m},\mathbf{g}\right) \right\vert ^{2}}{\sqrt{%
1-y^{2}\left( \theta _{\gamma }\right) }}\frac{\varepsilon _{1}\theta
_{\gamma }/m_{e}}{\left( g_{\perp }/m_{e}\right) \left[ g_{m\parallel
}/\delta -1-g_{\perp }^{2}/\left( 2\epsilon _{1}\delta \right) \right] ^{%
\frac{1}{2}}}.  \label{sigsumgxgy26}
\end{eqnarray}

In the second case, we assume that $\delta \sim 2\pi \theta \alpha /a_{1}$.
Now the main contribution into the sum in Eq. (\ref{sigcasegz0}) comes from
terms with $g_{y}=0$ and summations remain over $m$ and $n_{1}$, $g_{x}=2\pi
n_{1}/a_{1}$. The formula for the cross-section takes the form%
\begin{eqnarray}
\frac{d^{2}\sigma _{b}^{c}}{d\omega d\left( \epsilon _{1}\theta _{\gamma
}/m_{e}\right) } &\approx &\frac{e^{2}N}{\pi ^{2}\epsilon
_{1}^{2}N_{0}\Delta }\sum_{m,n_{1}}\frac{g_{m\perp }^{2}}{g_{m\parallel }^{2}%
}\left[ 1+\frac{\omega ^{2}}{2\epsilon _{1}\epsilon _{2}}-4y^{2}\left(
\theta _{\gamma }\right) \frac{\delta }{g_{m\parallel }}\left( 1-\frac{%
\delta }{g_{m\parallel }}\right) \right]  \notag \\
&&\frac{\left\vert F_{m}\left( \mathbf{g}_{m}\mathbf{u}_{0}\right)
\right\vert ^{2}\left\vert S\left( \mathbf{g}_{m},\mathbf{g}\right)
\right\vert ^{2}}{\sqrt{1-y^{2}\left( \theta _{\gamma }\right) }}\frac{%
\epsilon _{1}\theta _{\gamma }/m_{e}}{\left( g_{m\perp }/m_{e}\right) \left[
g_{m\parallel }/\delta -1-g_{m\perp }^{2}/\left( 2\epsilon _{1}\delta
\right) \right] ^{\frac{1}{2}}},  \label{sigsumg127}
\end{eqnarray}%
where
\begin{equation}
g_{m\parallel }\approx -mk_{z}+g_{x}\psi ,\quad \psi =\alpha \theta ,
\label{gmparpsi}
\end{equation}%
and the summation goes over the values $m$ and $n_{1}$ satisfying the
condition $g_{m\parallel }\geq \delta +g_{x}^{2}/\left( 2\varepsilon
_{1}\right) $.

We have carried out numerical calculations for the coherent bremsstrahlung
cross-section for various values of parameters in the case of ${\mathrm{SiO}}%
_{2}$ single crystal at zero temperature. To deal with an orthogonal
lattice, we choose as an elementary cell the cell including 6 atoms of
silicon and 12 atoms of oxygen (Shrauf elementary cell \cite{DanaDana}). For
this choice the $y$ and $z$ axes of the orthogonal coordinate system $\left(
x,y,z\right) $ coincide with the standard $Y$ and $Z$ -axes of the quartz
crystal, whereas the angle between the axes $x$ and $X$ is equal to $\pi /6$%
. For the potentials of atoms we take Moliere parametrization with%
\begin{equation}
u_{\mathbf{q}}^{\left( j\right) }=\sum_{i=1}^{3}\frac{4\pi Z_{j}e^{2}\alpha
_{i}}{q^{2}+\left( \chi _{i}/R_{j}\right) ^{2}}  \label{Molierpot}
\end{equation}%
where $\alpha _{i}=\left\{ 0.1,0.55,0.35\right\} ,$ $\chi _{i}=\left\{
6.0,1.2,0.3\right\} ,$ $R_{j}$ is the screening radius for the $j$-th atom
in the elementary cell.
\begin{figure}[tbph]
\begin{center}
\begin{tabular}{ccc}
\epsfig{figure=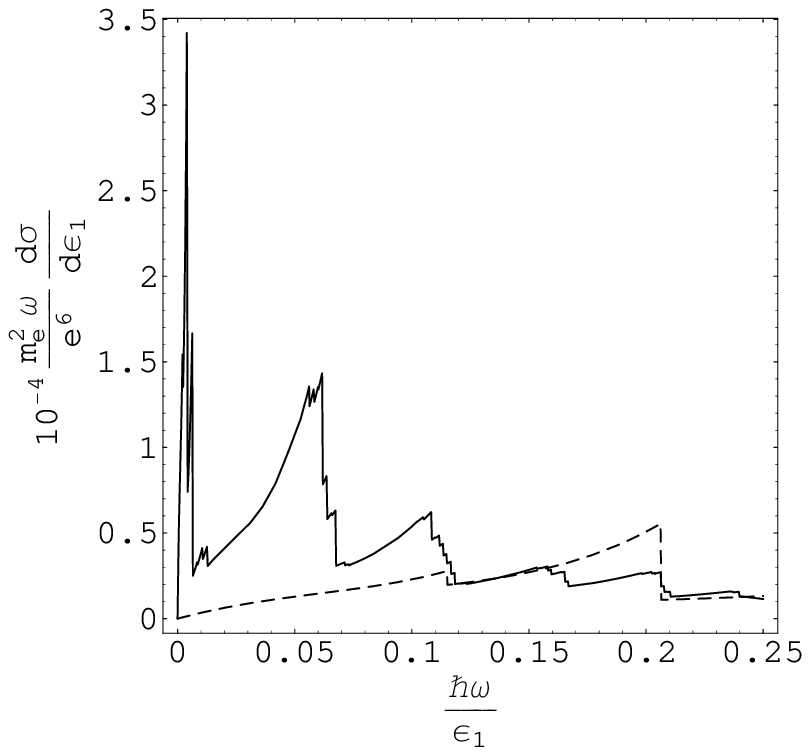,width=6cm,height=6cm} & \hspace*{0.5cm} & %
\epsfig{figure=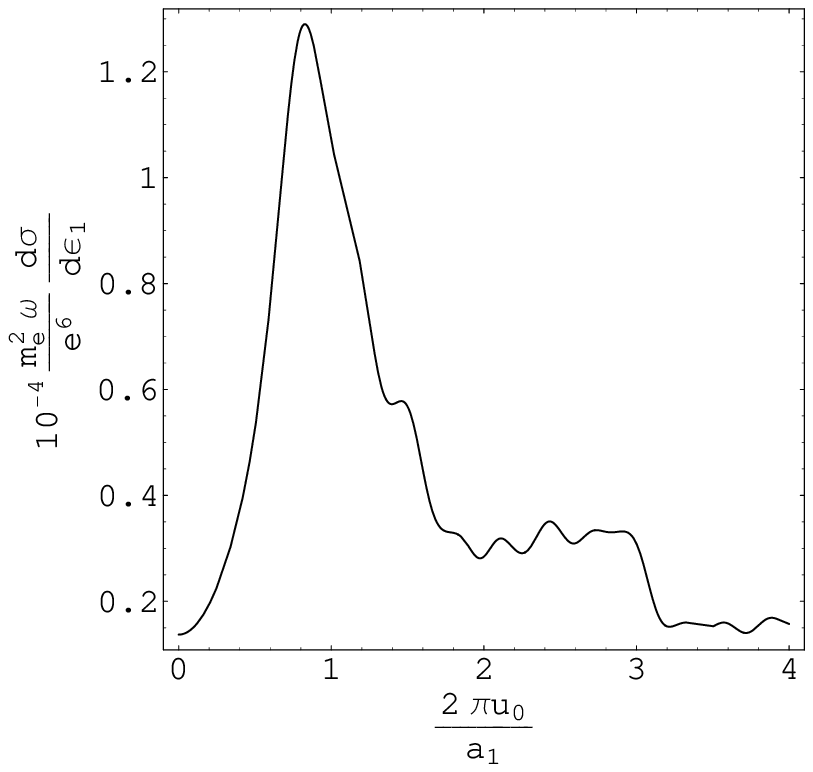,width=6cm,height=6cm}%
\end{tabular}%
\end{center}
\caption{Coherent bremsstrahlung cross-section, $10^{-4}(m_{e}^{2}\protect%
\omega /e^{6})d\protect\sigma _{b}^{c}/d\protect\omega $, evaluated by the
formula from ref. \protect\cite{Parazbrem}, as a function of $\protect\omega %
/\protect\epsilon _{1}$ \ for $2\protect\pi u_{0}/a_{1}=0$ (dashed curve), $2%
\protect\pi u_{0}/a_{1}=0.82$ (full curve), and as function of $2\protect\pi %
u_{0}/a_{1}$ (right panel) for the photon energy corresponding to $\protect%
\omega /\protect\epsilon _{1}=0.055$. The values for the other parameters
are as follows: $\protect\psi =0.00037,$ $\protect\epsilon _{1}=20GeV,$ $%
\protect\nu _{s}=5\cdot 10^{9}$ Hz.}
\label{brem20}
\end{figure}

The calculations are carried out for the sinusoidal transversal acoustic
wave of the S-type (the corresponding parameters can be found in Ref. \cite%
{Shaskol}) for which the vector of the amplitude of the displacement is
directed along $X$ direction of quartz single crystal, $\mathbf{u}%
_{0}=\left( u_{0},0,0\right) $, and the velocity is $4.687\cdot 10^{5}cm/$%
sec. The vector determining the direction of the hypersound propagation lies
in the plane $YZ$ and has the angle with the axis $Z$ equal to $0.295$ rad.
As the axis $z$ we choose the axis $Z$ of the quartz crystal. The
corresponding function $F\left( x\right) $ is determined by formula \ref%
{Fmdefin}. We have numerically evaluated the pair creation cross-section by
making use of formulae (\ref{sigsumg127}) for values of parameters $\epsilon
_{1},$ $\psi $, $u_{0}$ taken from \cite{Parazbrem} when one has an
enhancement of the cross-section.

\begin{figure}[tbph]
\begin{center}
\begin{tabular}{c}
\epsfig{figure=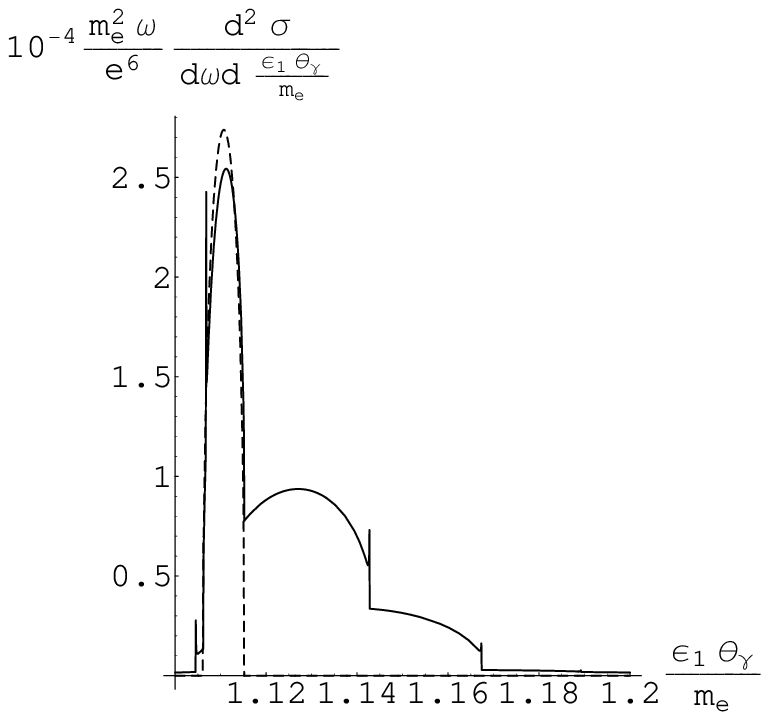,width=6cm,height=6cm}%
\end{tabular}%
\end{center}
\caption{Coherent bremsstrahlung cross-section, $10^{-4}(m_{e}^{2}\protect%
\epsilon _{1}/e^{6})d^{2}\protect\sigma _{b}^{c}/d\protect\omega d\left(
\protect\epsilon _{1}\protect\theta _{\protect\gamma }/m_{e}\right) $,
evaluated by formula ( \protect\ref{sigsumg127}), as a function of $\protect%
\omega \protect\theta _{\protect\gamma }/m_{e}$ for $2\protect\pi %
u_{0}/a_{1}=0$ (dashed curve), $2\protect\pi u_{0}/a_{1}=0.82$ (full curve),
$\protect\psi =0.00037$. The values for the other parameters are as follows:
$\protect\omega /\protect\epsilon _{1}=0.055,$ $\protect\epsilon _{1}=20GeV,$
$v_{s}=5\cdot 10^{9}$ Hz for the frequency of acoustic waves.}
\label{brem20dis}
\end{figure}

Numerical calculation show, that in dependence of the values for parameters,
the external excitation can either enhance or reduce the cross-section of
the bremsstrahlung process. As an illustration of the enhancement in the
cross-section integrated over the angle $\theta _{\gamma }$, on the left
panel of Fig. \ref{brem20} we have plotted the quantity $10^{-4}(m_{e}^{2}%
\omega /e^{6})d\sigma _{b}^{c}/d\omega $, evaluated by using the formula
from ref. \cite{Parazbrem}, as a function of the ratio $\omega /\epsilon
_{1} $ \ in the case of ${\mathrm{SiO}}_{2}$ mono crystal and Moliere
parametrization of the screened atomic potential for $2\pi u_{0}/a_{1}=0$
(dashed curve), $2\pi u_{0}/a_{1}=0.82$ (full curve). On the right panel the
same quantity is plotted as a function of $2\pi u_{0}/a_{1}$ for the photon
energy corresponding to $\omega /\epsilon _{1}=0.055$. The values for the
other parameters are taken as follows: $\epsilon _{1}=20$ GeV, $\psi
=0.00037 $, $\nu _{s}=5\cdot 10^{9}$ Hz for the frequency of acoustic waves.
For the amplitude of the deformation field corresponding to the numerical
data of Fig. \ref{brem20} the relative displacement of the neighboring atoms
is of the order $10^{-3}\mathring{A}$, which is much smaller than the
interatomic distance ($\sim 5\mathring{A}$). For these values of parameters,
when one has an enhancement of the cross-section integrated over the angle $%
\theta _{\gamma }$, we have numerically analyzed the angular dependence of
the bremsstrahlung cross-section by making use of formula (\ref{sigsumg127}%
). In Fig. \ref{brem20dis} the quantity $10^{-4}(m_{e}^{2}\omega
/e^{2})d^{2}\sigma _{b}^{c}/d\omega d\left( \epsilon _{1}\theta _{\gamma
}/m_{e}\right) $ is depicted as a function of $\epsilon _{1}\theta _{\gamma
}/m_{e}$ in the case of ${\mathrm{SiO}}_{2}$ monocrystal for $u_{0}=0$
(dashed curve) and $2\pi u_{0}/a_{1}=0.82$ (full curve). The values for the
other parameters are taken as follows: $\omega /\epsilon _{1}=0.055,$ $%
\epsilon _{1}=20$ GeV,$\ \nu _{s}=5\cdot 10^{9}$ Hz, $\psi =0.00037$.

\begin{figure}[tbph]
\begin{center}
\begin{tabular}{ccc}
\epsfig{figure=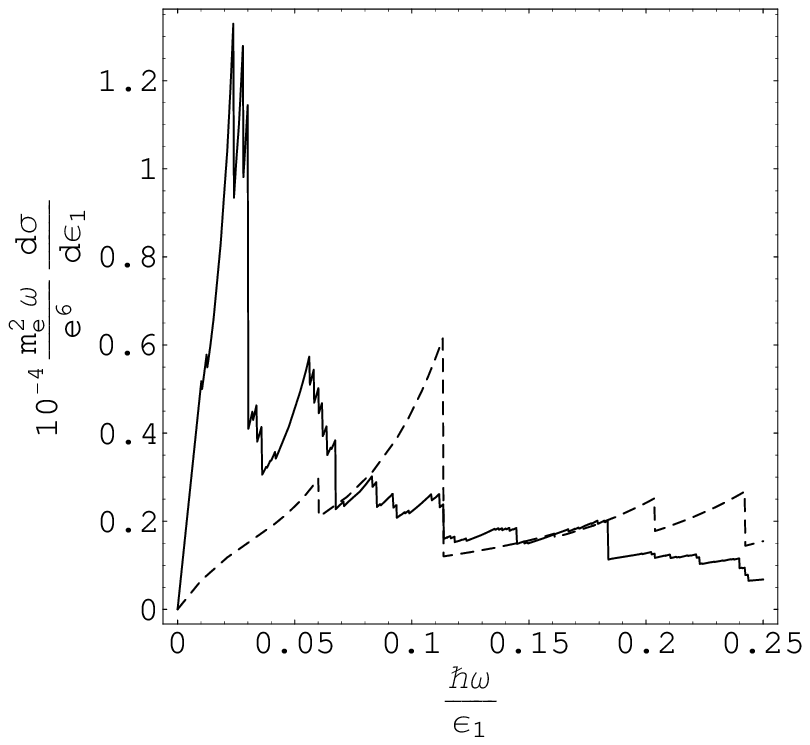,width=6cm,height=6cm} & \hspace*{0.5cm} & %
\epsfig{figure=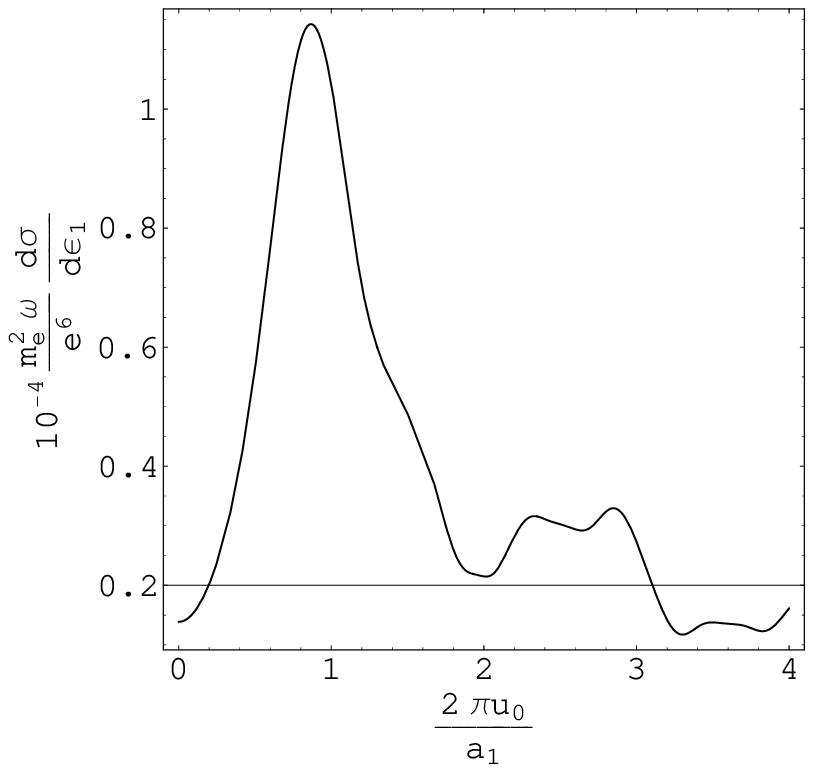,width=6cm,height=6cm}%
\end{tabular}%
\end{center}
\caption{Coherent bremsstrahlung cross-section, $10^{-4}(m_{e}^{2}\protect%
\omega /e^{6})d\protect\sigma _{b}^{c}/d\protect\omega $, evaluated by the
formula from ref. \protect\cite{Parazbrem}, as a function of $\protect\omega %
/\protect\epsilon _{1}$ \ for $2\protect\pi u_{0}/a_{1}=0$ (dashed curve), $2%
\protect\pi u_{0}/a_{1}=0.85$ (full curve), and as function of $2\protect\pi %
u_{0}/a_{1}$ (right panel) for the photon energy corresponding to $\protect%
\omega /\protect\epsilon _{1}=0.0266$. The values for the other parameters
are as follows: $\protect\psi =0.000364,$ $\protect\epsilon _{1}=10GeV,$ $%
\protect\nu _{s}=5\cdot 10^{9}$ Hz.}
\label{brem10}
\end{figure}

In order to see the dependence of the results on the energy of the incoming
electron, in Fig. \ref{brem10} we have presented the quantity $%
10^{-4}(m_{e}^{2}\omega /e^{6})d\sigma _{b}^{c}/d\omega $, evaluated by
using the formula from ref. \cite{Parazbrem}, as a function of the ratio $%
\omega /\epsilon _{1}$ \ in the case of ${\mathrm{SiO}}_{2}$ mono crystal
for $2\pi u_{0}/a_{1}=0$ (dashed curve), $2\pi u_{0}/a_{1}=0.85$ (full
curve). On the right panel the same quantity is plotted as a function of $%
2\pi u_{0}/a_{1}$ for the photon energy corresponding to $\omega /\epsilon
_{1}=0.0266$. The values for the other parameters are taken as follows: $%
\epsilon _{1}=10$ GeV, $\psi =0.000364$, $\nu _{s}=5\cdot 10^{9}$ Hz for the
frequency of acoustic waves.

\begin{figure}[tbph]
\begin{center}
\begin{tabular}{c}
\epsfig{figure=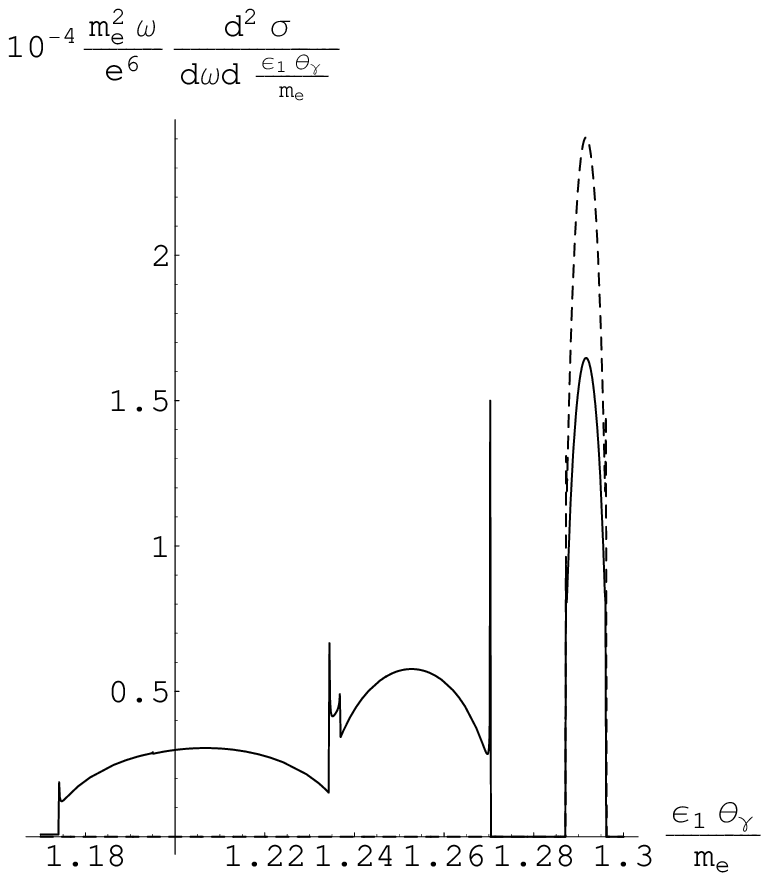,width=6cm,height=6cm}%
\end{tabular}%
\end{center}
\caption{Coherent bremsstrahlung cross-section, $10^{-4}(m_{e}^{2}\protect%
\epsilon _{1}/e^{6})d^{2}\protect\sigma _{b}^{c}/d\protect\omega d\left(
\protect\epsilon _{1}\protect\theta _{\protect\gamma }/m_{e}\right) $,
evaluated by formula ( \protect\ref{sigsumg127}), as a function of $\protect%
\omega \protect\theta _{\protect\gamma }/m_{e}$ for $2\protect\pi %
u_{0}/a_{1}=0$ (dashed curve), $2\protect\pi u_{0}/a_{1}=0.85$ (full curve),
$\protect\psi =0.000364$. The values for the other parameters are as
follows: $\protect\omega /\protect\epsilon _{1}=0.0266,$ $\protect\epsilon %
_{1}=10GeV,$ $v_{s}=5\cdot 10^{9}$ Hz for the frequency of acoustic waves.}
\label{brem10dis}
\end{figure}

In Fig.\ref{brem10dis} we have depicted the quantity $10^{-4}(m_{e}^{2}%
\omega /e^{2})d^{2}\sigma _{b}^{c}/d\omega d\left( \epsilon _{1}\theta
_{\gamma }/m_{e}\right) $ as a function of $\epsilon _{1}\theta _{\gamma
}/m_{e}$ in the case of ${\mathrm{SiO}}_{2}$ mono crystal for $u_{0}=0$
(dashed curves), $2\pi u_{0}/a_{1}=0.85$ and for $\psi =0.000364$. The
values for the other parameters are as follows: $\omega /\epsilon
_{1}=0.0266,$ $\epsilon _{1}=10$ GeV, $v_{s}=5\cdot 10^{9}$ Hz for the
frequency of acoustic waves.

As we see from the presented example, the presence of the deformation field
leads to the appearance of additional peaks in the angular distribution of
the emitted photon even for such ranges of values of an angle of photon
momentum, where due to the requirement $-1\leq y\leq 1$ the cross-section of
process is zero when the deformation is absent. As we have already mentioned
before, this is related to that in the presence of the deformation field the
number of possibilities to satisfy the condition $g_{m\parallel }\geq \delta
+g_{m\perp }^{2}/\left( 2\epsilon _{1}\right) $ in the summation in formula (%
\ref{sigcohgeneral}) increases due to the presence of the additional term $%
mk_{s\parallel }$ in the expression for $g_{m\parallel }$.
\newline

\section{Conclusion}

\label{sec4:conc}

The present paper is devoted to the investigation of the angular
distribution of the electron in the bremsstrahlung process by high-energy
electrons in a crystal with a complex lattice base in the presence of
deformation field of an arbitrary periodic profile. The latter can be
induced, for example, by acoustic waves. The influence of the deformation
field can serve as a possible mechanism to control the angular-energetic
characteristics of the created particles. The importance of this is
motivated by that the coherent bremsstrahlung of high-energy electrons
moving in a crystal is one of the most effective methods to produce intense
beams of highly polarized and monochromatic photons. In a crystal the
cross-section is a sum of coherent and incoherent parts. The coherent part
of the cross-section per single atom, averaged on thermal fluctuations, is
given by formula (\ref{sigcohgeneral}). In this formula the factor $%
\left\vert F_{m}\left( \mathbf{g}_{m}\mathbf{u}_{0}\right) \right\vert ^{2}$
is determined by the function describing the displacement of the atoms due
to the deformation field, and the factor $\left\vert S\left( \mathbf{g}_{m},%
\mathbf{g}\right) \right\vert ^{2}$ is determined by the structure of the
crystal elementary cell. Compared with the cross-section in an undeformed
crystal, formula (\ref{sigcohgeneral}) contains an additional summation over
the reciprocal lattice vector of the one-dimensional superlattice induced by
the deformation field. We have argued that the influence of the deformation
field on the cross-section can be remarkable under the condition $4\pi
^{2}u_{0}/a\gtrsim \lambda _{s}/l_{c}$. Note that for the deformation with $%
4\pi ^{2}u_{0}/a>1$ this condition is less restrictive than the naively
expected one $\lambda _{s}\leq l_{c}$. The role of coherence effects in the
pair creation cross-section is essential when the photon enters into the
crystal at small angles with respect to a crystallographic axis. In this
case the main contribution into the coherent part of the cross-section comes
from the crystallographic planes, parallel to the chosen axis (axis $z$ in
our consideration). The behavior of this cross-section as a function on the
positron energy essentially depends on the angle $\theta $ between the
projection of the electron momentum on the plane $\left( x,y\right) $ and $y$%
-axis. When the electron enters into the crystal near a crystallographic
plane, two cases have to be distinguished. For the first one $\theta \sim
a_{2}/2\pi l_{c}$ the formula (\ref{sigcasegz0}) is further simplified to
the form (\ref{sigsumgxgy26}) under the assumption $\mathbf{u}_{0}\perp
\mathbf{a}_{1}$. In the second case one has $\psi =\alpha \theta \sim
a_{1}/2\pi l_{c}$, and the main contribution into the cross-section comes
from the crystallographic planes parallel to the incidence plane. The
corresponding formula for the cross-section takes the form (\ref{sigsumg127}%
). The numerical calculations for the cross-section are carried out in the
case of SiO$_{2}$ single crystal with the Moliere parametrization of the
screened atomic potentials and for the deformation field generated by the
transversal acoustic wave of $S$ - type with frequency 5 GHz. Examples of
numerical results are depicted in figures. The numerical calculations for
values of the parameters in the problem when one has an enhancement of the
cross-section show that the presence of the deformation field leads to the
appearance of additional peaks in the angular distribution of the radiated
photon even for such ranges of values of an angle of a photon, where due to
the requirement $-1\leq y\leq 1$ the cross-section is zero when deformation
is absent. This can be used to control the parameters of the photon sources.

\section*{Acknowledgment}

I am grateful to Aram Saharian for valuable discussions and suggestions.

\end{document}